\documentclass[%
 reprint,%
 amssymb, amsmath,%
 aip,cha,%
]{revtex4-1}

\usepackage{graphics}
\usepackage{amsmath}
\usepackage{amsfonts}
\usepackage{latexsym}
\usepackage{amsbsy}
\usepackage{epstopdf}
\usepackage{bm}%

\usepackage{color}

\usepackage{ulem} 

\newcommand{\dyadic}[1]{{#1}
\setbox0=\hbox{$\scriptstyle\leftrightarrow$}
   \setbox2=\hbox{$#1$}
   \dimen0=.5\wd0 \advance\dimen0 by-.5\wd2
   \advance\dimen0 by-\wd0
   \kern\dimen0
{^{\hbox{$\scriptstyle\leftrightarrow$}}}}

\begin{document}

\title{
Detecting and Receiving Phase Modulated Signals with a Rydberg Atom-Based Mixer
}
\thanks{Publication of the U.S. government, not subject to U.S. copyright.}
\author{Christopher~L.~Holloway}
\email{christopher.holloway@nist.gov}
\author{Matthew T. Simons}
\author{Joshua A. Gordon}
\author{David Novotny}
\affiliation{National Institute of Standards and Technology, Boulder,~CO~80305, USA}

\date{\today}

\begin{abstract}

Recently, we introduced a Rydberg-atom based mixer capable of detecting and measuring the phase of a radio-frequency field through the electromagnetically induced transparency (EIT) and  Autler-Townes (AT) effect. The ability to measure phase with this mixer allows for an atom-based receiver to detect digital modulated communication signals. In this paper, we demonstrate detection and reception of digital modulated signals based on various phase-shift keying approaches. We demonstrate Rydberg atom-based digital reception of binary phase-shift keying (BPSK), quadrature phase-shift keying (QPSK), and quadrature amplitude (QAM) modulated signals over a 19.626~GHz carrier to transmit and receive a bit stream in cesium vapor. We present measured values of Error Vector Magnitude (EVM, a common communication metric used to assess how accurate a symbol or bit stream is received) as a function of symbol rate for BPSK, QPSK, 16QAM, 32QAM, and 64QAM modulation schemes.  These results allow us to discuss the bandwidth of a Rydberg-atom based receiver system.
\end{abstract}

\maketitle

\begin{figure}
\centering
\scalebox{.4}{\includegraphics*{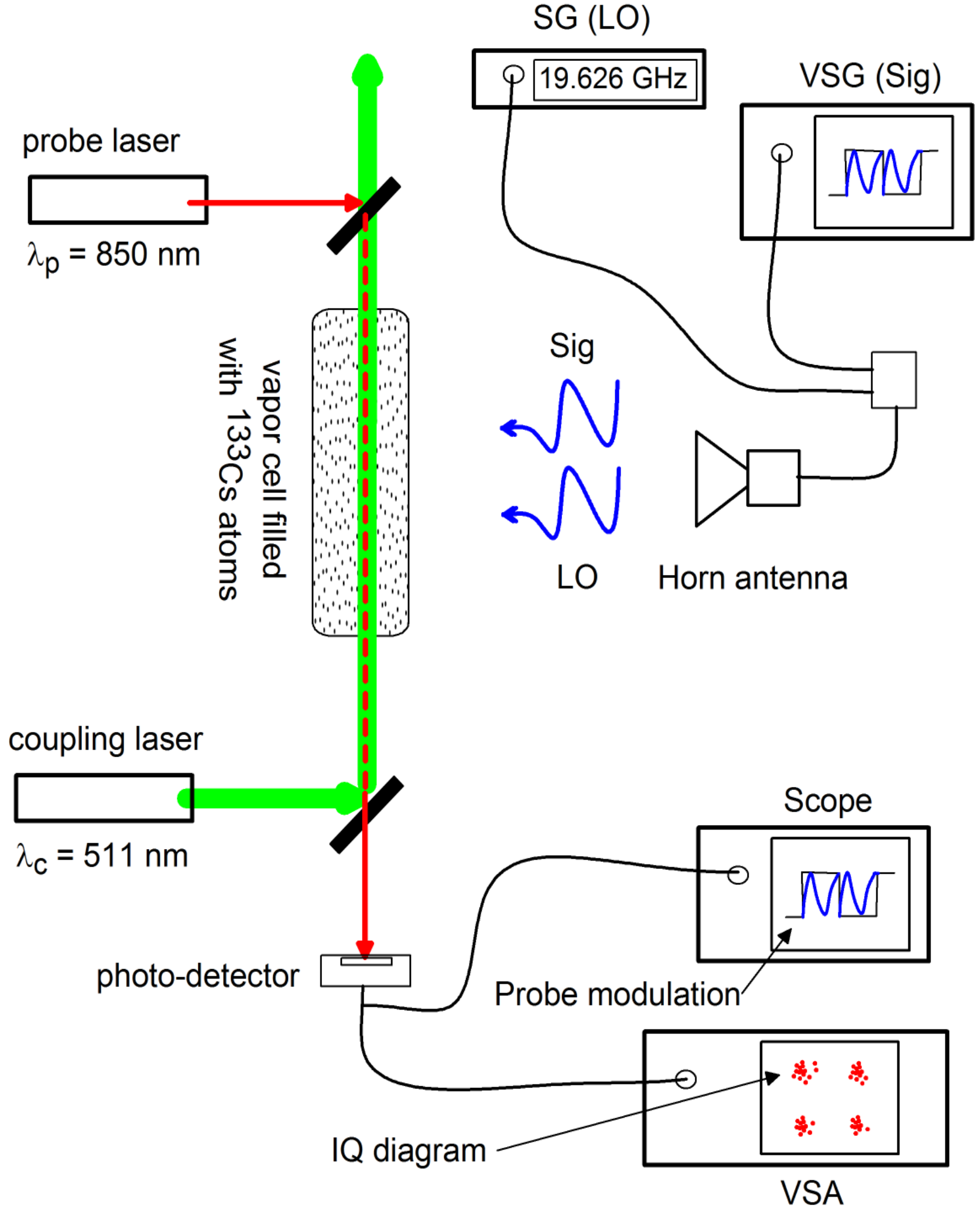}}
\caption{Experimental setup for receiving digital modulated signals.}
\label{f1}
\end{figure}

The use of Rydberg states of an alkali atomic vapor placed in glass cells for the development of radio frequency (RF) electric (E) field strength and power metrology techniques has made great strides in recent years \cite{r2, r3, r4, r5, r6, r7, r8, r9, r10, r11, r12, r13, r14, r15, r16, r17, r18, r19, r20}.  Electromagnetically-induced transparency (EIT) is used in this approach for E-field sensing, performed either when an RF field is on-resonance of a Rydberg transition (using Autler-Townes (AT) splitting) or off-resonance (using AC Stark shifts).  This Rydberg-atom based sensor can act as compact quantum-based reciever/antenna for communication applications to detect and receive modulated signals. The idea of a Rydberg receiver/antenna for modulated signals was demonstrated in Ref.\cite{dan}, and further developments are found in Refs.\cite{rc1, rc2, amfmstereo, biterror, rc3, rc4}. 
Most of these demonstrations are limited to amplitude modulation (AM) and/or frequency modulation (FM) schemes. However, reception of one form of digital signals has been demonstrated by amplitude modulating a carrier\cite{rc1}. This EIT scheme has been very successful in detecting the amplitude of continuous-wave (CW) carriers (which is all that is required for AM or FM signals\cite{amfmstereo, rc3}). In order to detect and receive phase modulated signals [the basis of  most digital modulation schemes, e.g., binary phase-shift keying (BPSK), quadrature phase-shift keying (QPSK), and quadrature amplitude modulation (QAM)], the phase of an RF signal is required.  However, the ability to measure the phase of an RF  signal with Rydberg atoms has not been possible until recently\cite{phase}. By using a Rydberg atom-based mixer (depicted in Fig.~\ref{f1}) that was recently developed \cite{phase, weak}, we demonstrate the ability to detect a phase modulated RF carrier and in turn, detect and receive BPSK, QPSK, and QAM signals.

A widely used modulation scheme for digital communications is phase-shift keying (PSK) using both binary and quadrature PSK (BPSK and QPSK)\cite{qpsk}. In these modulation schemes, data is transmitted by changing (or modulating) the phase of CW carrier. BPSK uses two different phase states to transmit data, in which the carrier frequency phase is changed between 0$^o$ and 180$^o$. Each phase state represents one transmitted symbol and each symbol is mapped into bits ``1'' or ``0''.  QPSK is a type of PSK where each transmitted symbol (or phase state) is mapped into two bits. This is done by choosing one of four possible phases applied to a CW carrier [e.g., $45^o$ (binary state ``00'') , $135^o$ (binary state ``01''), -$45^o$ (binary state ``10''), and -$135^o$ (binary state ``11'')].  Using both phase and amplitude, this idea is extended to quadrature amplitude modulation (QAM), where 16QAM corresponds to 16 phase and amplitude states, each phase state is a transmitted symbol (each symbol corresponds to 4 bits, ``0000'', ``1000'', ``1100'', etc..), 32QAM corresponds to 32 phase and amplitude states, each phase state is a transmitted symbol (each symbol corresponds to 5 bits), 64QAM corresponds to 64 phase and amplitude states, each phase state is a transmitted symbol (each symbol corresponds to 6 bits). Thus, to receive BPSK, QPSK and QAM signals, one needs to measure and detect the phase and amplitude of the CW carrier. The Rydberg-atom based mixer\cite{phase} allows us to measure the phase and amplitude of a carrier and we use this mixer to receive BPSK, QPSK, 16QAM, 32QAM, and 64QAM modulated signals.

Details of how the atom-based mixer works are given in Ref.\cite{phase} and we give a brief discussion here. An RF field (labeled ``LO'' in Fig.~\ref{f1}) on-resonance with the Rydberg transition acts as a local oscillator (LO). The ``LO'' field causes the EIT/AT effect in the Rydberg atoms which is used to down-convert a second, co-polarized RF field. This second field is detuned from the ``LO'' field and is the digital modulated carrier (labeled ``SIG'' in Fig~\ref{f1}). The frequency difference between the LO and the SIG is an intermediate frequency (IF) and the IF is detected by optically probing the Rydberg atoms (Fig.~\ref{f1}). The phase of the IF signal corresponds directly to the relative phase between the ``LO'' and ``SIG'' signals.   In effect, the atom-based mixer does all the down-conversion of the ``SIG'', and a direct read-out of the phase of SIG is obtained by the probe laser propagating through the atomic vapor. By measuring the relative phase shift of the IF signal (via a photodetector) we can determine the phase states of BPSK, QPSK, and QAM signals.

The EIT/AT technique involves monitoring the transmission of a ``probe'' laser through the vapor cell. A second laser (``coupling'' laser) establishes a coherence in the atomic states, and enhances the probe transmission through the atoms. An applied RF field (the LO field in our case) alters the susceptibility of the atomic vapor, which results in a change in the probe laser transmission. As shown in Ref.\cite{phase}, the presence of both LO and SIG field creates a beat-note, and the beat-note results amplitude modulation (AM) of the probe transmission, where the amplitude of the probe transmission varies as $\cos(2\pi f_{IF}t+\Delta\phi)$ (where $f_{IF}$ is the frequency of the IF field and $\Delta\phi$ is the phase difference between the LO and SIG field). This AM of the probe laser transmission can be detected with a photodetector and used to determine the phase of the SIG signal. For a pure AM or FM carrier, the Rydberg atoms automatically demodulate the carrier and output of the photodetector gives a direct read-out of the bassband signal (the information being transmitted). For a phase modulated carrier, the Rydberg atoms automatically down-convert the carrier to the IF, which contains the phase states of the different phase modulation schemes.

A diagram of the experimental setup is shown in Fig.~\ref{f1}.  To generate EIT we use cesium ($^{133}$Cs) atoms.  The probe laser is tuned to the ${\rm D}_2$ transition for $^{133}$Cs ($6S_{1/2}$-$6P_{3/2}$ or wavelength of $\lambda_p=850.53~nm$) focused to a full-width at half maximum (FWHM) of 425~$\mu$m, with a power of 41.2~$\mu$W. To produce an EIT signal, we couple to the $^{133}$Cs $6P_{3/2}$-$34D_{5/2}$ states by applying a counter-propagating coupling laser at $\lambda_c=511.1480$~nm with a power of 48.7~mW, focused to a FWHM of 620~$\mu$m. We use a signal generator (SG) to apply an LO field at 19.626~GHz to couple states $34D_{5/2}$ and $35P_{3/2}$. While we use 19.626~GHz in these experiments, this approach can work at carriers from 100~MHz to 1~THz (because of the broadband nature of the EIT/AT appraoch\cite{r2, r3}).  To generate the modulated SIG field we use a vector signal generator (VSG). The VSG applies a given digital modulation to a CW carrier. We set the frequency of the CW SIG field to 19.626~GHz+$f_{IF}$ (where the $f_{IF}$ is changed during these experiments). The output from  the SG and the VSG were connected to a standard gain horn antenna via a power combiner.   The output of the photodetector was connected to the input of a vector signal analyzer (VSA). The Rydberg atoms automatically down-converts the modulated carrier to the IF (the amplitude of the probe laser transmission) and the signal analyzer can detect the phase change of the IF signal and hence detect the phase state of the signal.  The output of the photodetector was also sent to an oscilloscope.

\begin{figure}
\centering
\scalebox{.3}{\includegraphics*{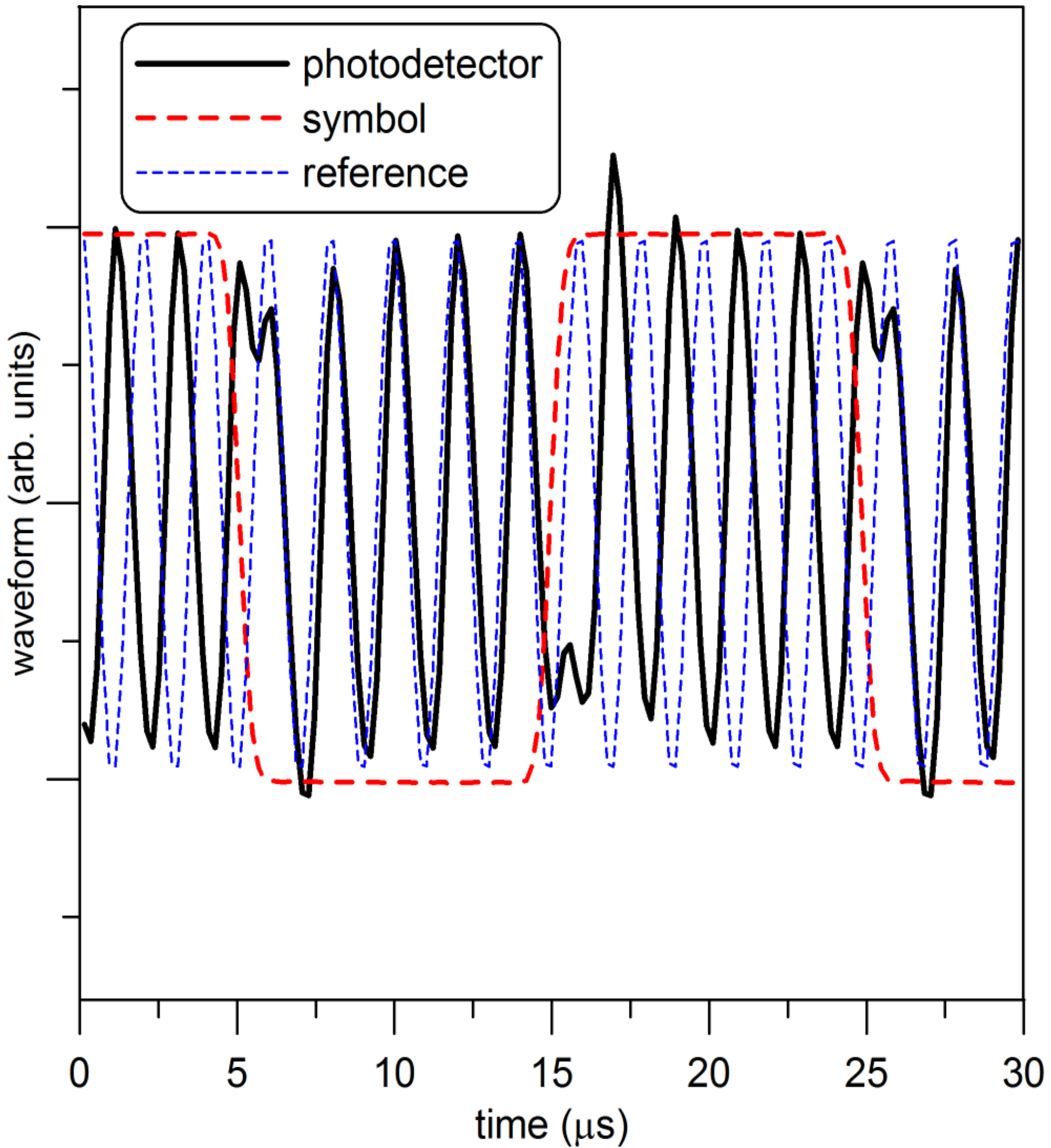}}
\caption{Signal detected on the photodetector (measured on the oscilloscope) for BPSK modulation for $f_{IF}=500$~kHz and symbol rate of 100~kSym/s (symbol period of 10~$\mu$s.)}
\label{bpsk}
\end{figure}

We first demonstrate the ability to receive a BPSK signal. Fig.~\ref{bpsk} shows the signal detected on the photodetector (measured on the oscilloscope) for a BPSK modulation for $f_{IF}=500$~kHz and symbol period of 1~$\mu$s (i.e., a symbol rate of 1kSym/s or 1 kbit/s). Also on the figure is a reference signal. Comparing the reference signal with the photodetector signal shows the phase shift in the signal when the symbol state changes (represented by the square wave in the figure).

\begin{figure}
\centering
\scalebox{.5}{\includegraphics*{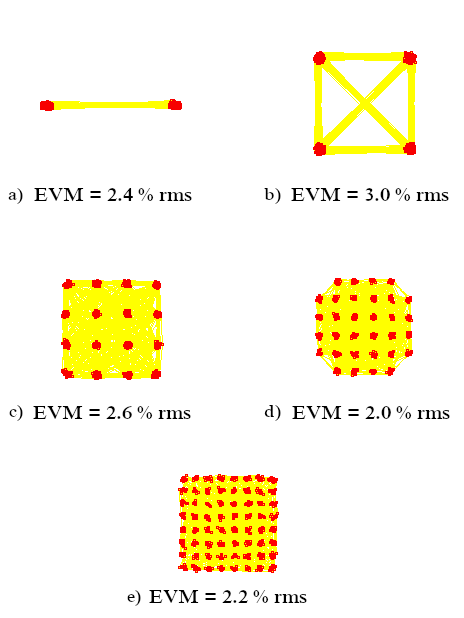}}
\caption{Measured IQ digrams: (a) BPSK, (b) QPSK, (c) 16QAM, (d) 32QAM, and (e) 64QAM. The EVM for each case is indicted as well. The bandwidth of both the photodetector and the VSA where 10~MHz.}
\label{IQplot}
\end{figure}

In communications, an IQ constellation diagram (IQ stands for in-phase and quadrature components of the modulated signal: also called a polar or vector diagram) is typically used to represent the phase state of a symbol (i.e., in our case the phase and amplitude of the IF signal).   Furthermore, a metric to assess how well a digital signal (a bit stream) is detected is the error vector magnitude (EVM)\cite{evm}. EVM is an error vector of the measured (received) phase/amplitude state compared to the ideal state and is basically an assessment of the received modulation quality. The VSA can generate the IQ diagram for the detected signal and calculate the EVM of the received bit stream. The IQ diagram for receiving 2047 symbols is shown in Fig.~\ref{IQplot}. Fig.~\ref{IQplot} shows the received IQ diagrams for the Rydberg-atom recevier for five different modulation schemes (BPSK, QPSK, 16QAM, 32QAM, and 64QAM), each with an IF=1~MHz and symbol rate of 100~kSym/s.

\begin{figure}
\centering
\scalebox{.2}{\includegraphics*{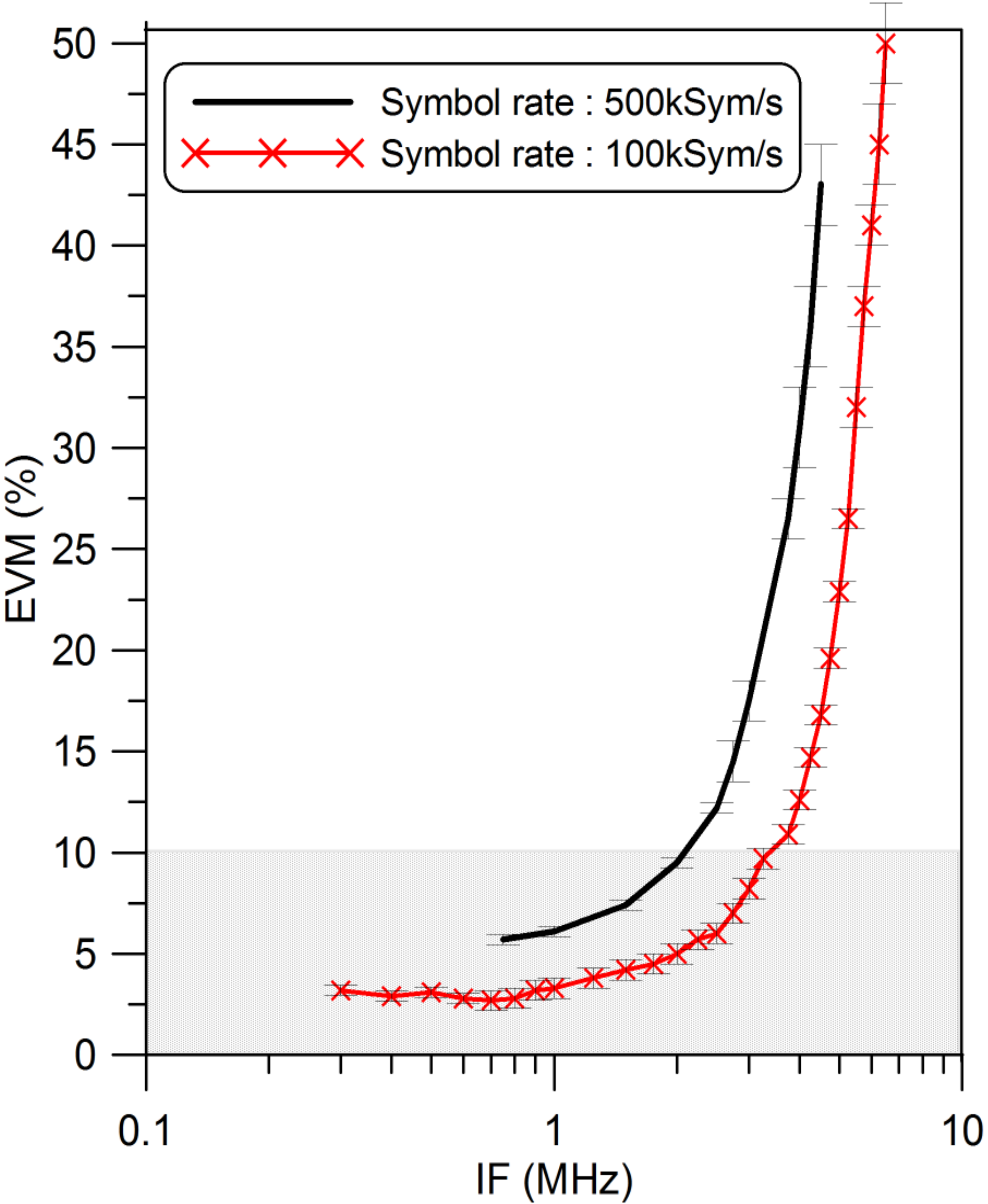}}\\
\caption{Measured EVM for BPSK for different IF. The error bars represent the variability in the measured EVM. The bandwidth of both the photodetector and the VSA where 10~MHz.}
\label{evmif}
\end{figure}

\begin{figure}
\centering
\scalebox{.2}{\includegraphics*{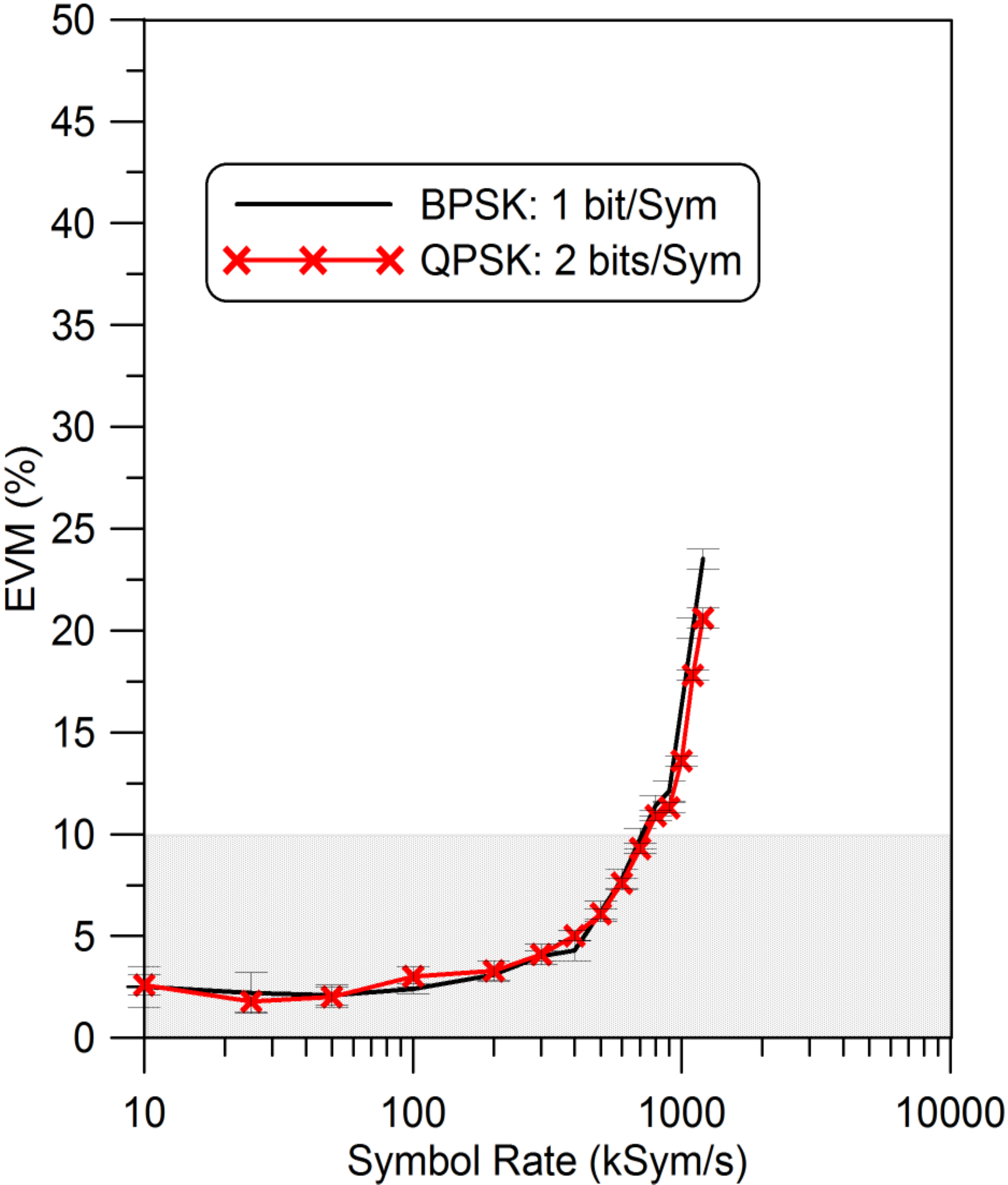}}\hspace{2mm}
\scalebox{.2}{\includegraphics*{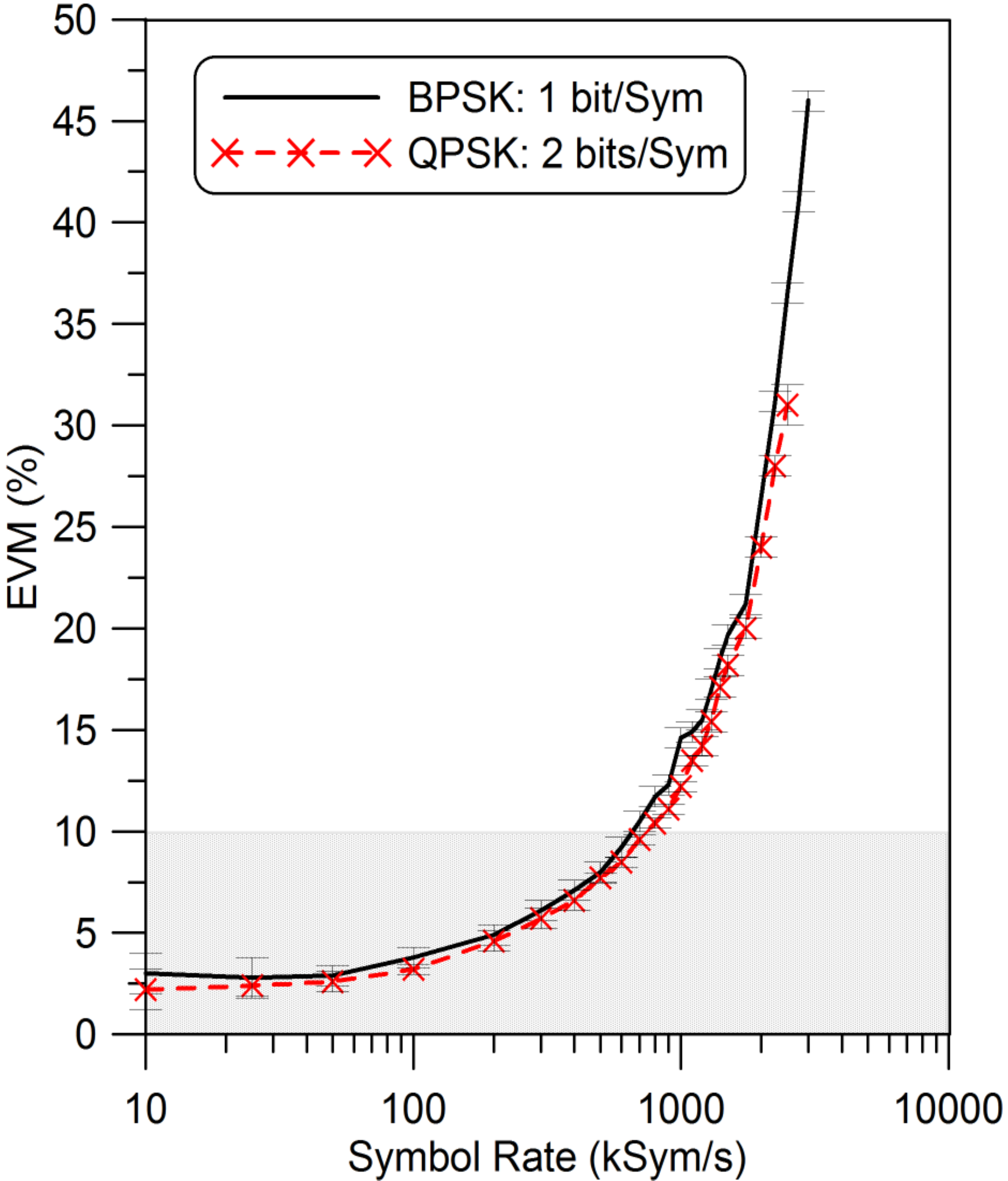}}\\
\begin{centering}
{(a) IF=1~MHz \hspace{20mm} (b) IF=2~MHz}\\
\end{centering}
\caption{Measured EVM for BPSK and QPSK: (a) IF=1~MHz and (b) IF=2~MHz. The bandwidth of both the photodetector and the VSA where 10~MHz.}
\label{evmall}
\end{figure}

We first looked at the bandwidth of the Rydberg atom-based receiver. This bandwidth limit is due to the time required to populate the atoms to a Rydberg state. A numerical time-domain calculation of the master equation for the density matrix components given in Ref.\cite{r15} shows that the population of the Rydberg state reaches steady-state around 1~$\mu$s, but has significant population by 0.1~$\mu$s to 0.3~$\mu$s, which implies the atoms can respond on the order of 3~MHz to 10~MHz. As we will see, while the Rydberg state may not be fully populated in 0.3~$\mu$s (3~MHz), the atom-based mixer can detect and receive digital signals for data-rate above 5~MHz (but the EVM starts to become large). For this atom-based mixer approach, varying the IF value gives an indication of the maximum data-rate for digital signals that can be detected.  In effect, the atoms respond to the IF signal, as a result, the higher the IF the faster the atoms have to respond. Fig.~\ref{evmif} shows the EVM as a function of IF for a BPSK signal for two different symbols rates.  We see that around 1~MHz, the EVM starts to increases, and around 2~MHz to 3~MHz, the EVM increases above 10~$\%$, but data is still received for IF$>3~$MHz.  Next, we set IF to 1~MHz and 2~MHz and varied the symbol rate. Fig.~\ref{evmall} show the EVM as a function of symbol rate for BPSK.  Here we see that that the EVM is below 5~$\%$ for symbol rates below 400~kSym/s for both IF values. The EVM approaches 10~$\%$ for symbol rate around 700~kSym/s in both cases. The EVM continues to increase with increasing symbol rate.
We should point out, that as one might expect, once the period of the IF becomes smaller than the symbol period it becomes difficult to detect the different phases of the carrier (i.e., when the IF wavelength is larger than the symbol length). While the high symbol rates are approaching the bandwidth of the Rydberg-atoms, the atom-base mixer still detects and receives BPSK signals with the caveat that the EVM does increase with high symbol rate.


Next, we transmitted a QPSK signal (an example of an IQ diagram is shown in Fig.~\ref{IQplot}).  The EVM for QPSK versus symbol rate is shown in Fig.~\ref{evmall}. We see that the QPSK follows the BPSK results. However, keep in mind the QPSK transmits 2~bits/symbol while BPSK transmits only 1~bit/symbol. Here again, once the period of the IF becomes smaller than the symbol rate, it becomes difficult to detect the phase states of the carrier.

\begin{figure}
\centering
\scalebox{.2}{\includegraphics*{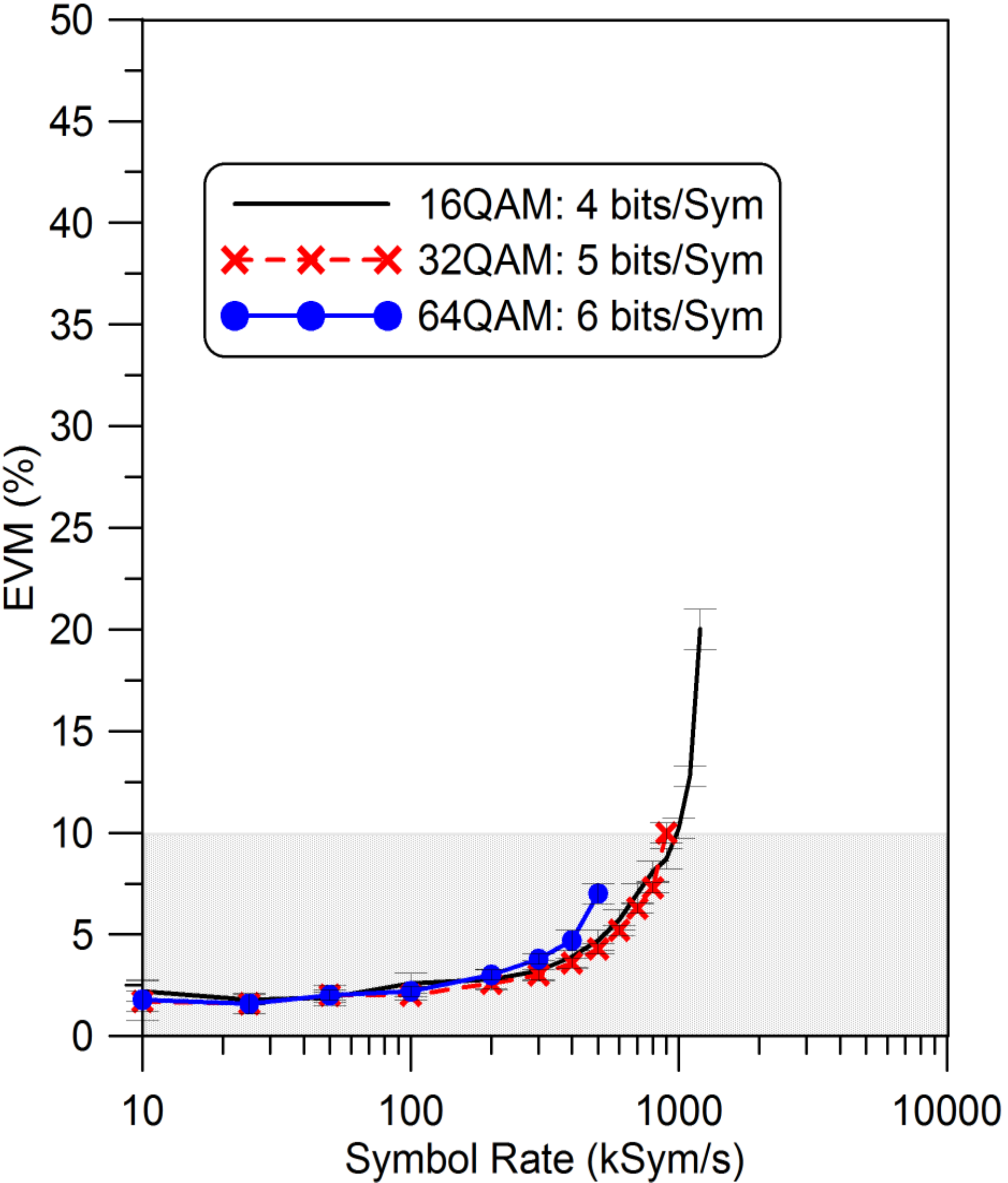}}\hspace{2mm}
\scalebox{.2}{\includegraphics*{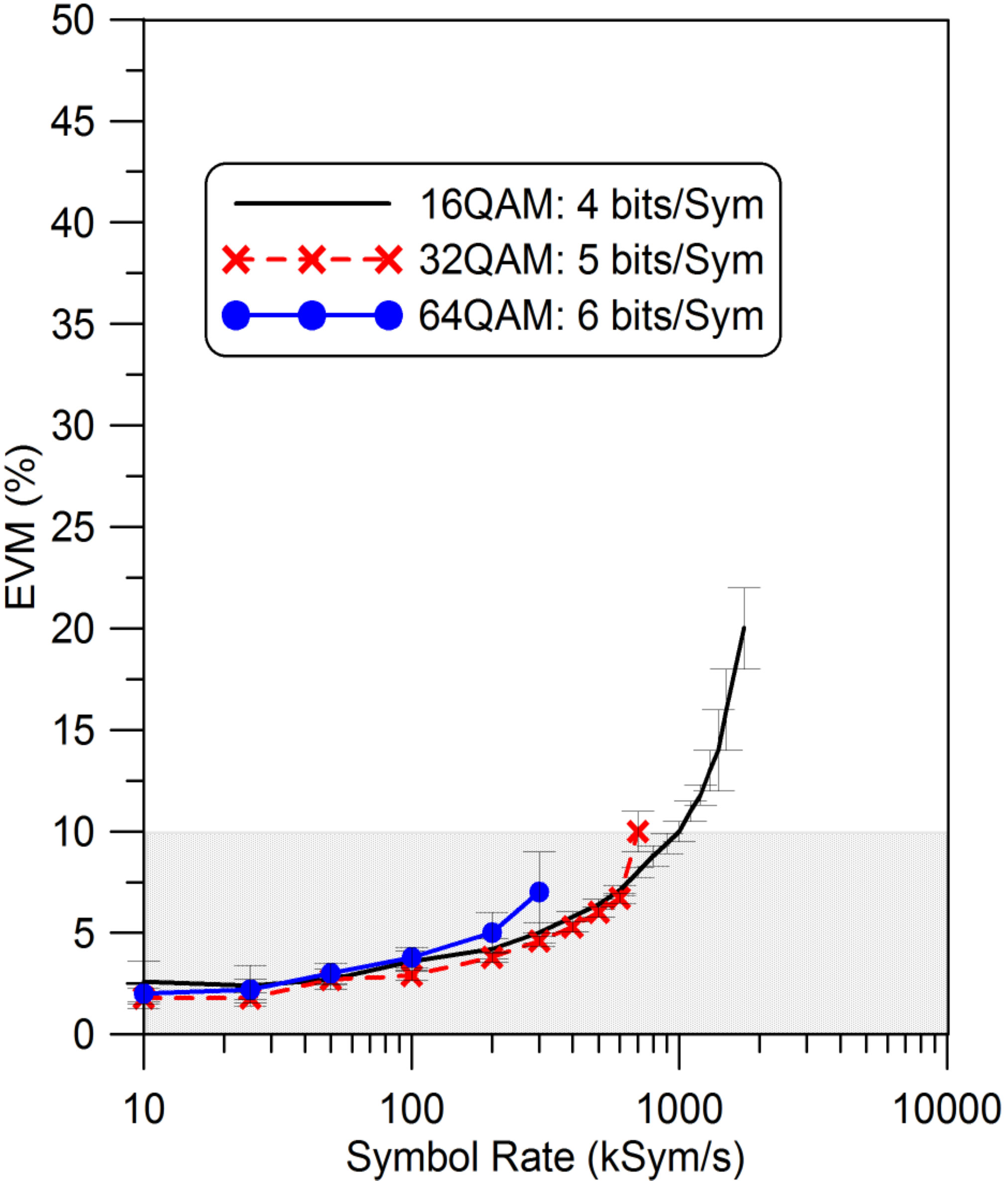}}\\
\begin{centering}
{(a) IF=1~MHz \hspace{20mm} (b) IF=2~MHz}\\
\end{centering}
\caption{Measured EVM for 16QAM, 32QAM, and 64QAM: (a) IF=1~MHz and (b) IF=2~MHz. The bandwidth of both the photodetector and the VSA where 10~MHz.}
\label{evmqam}
\end{figure}

Finally, we transmitted 16QAM, 32QAM, and 64QAM signals (IQ diagrams are shown in Fig.~\ref{IQplot}). These 16QAM, 32QAM, and 64QAM are actually transmitting 4~bits/symbol, 5~bits/symbol, and 6~bits/sybmol, respectively. The EVM for 16QAM, 32QAM, and 64QAM are shown in Fig.~\ref{evmqam}. From the IQ diagrams, we see that the phase states for the various QAM schemes become more crowded as the number of bits per symbol increases (i.e., going from 16QAM to 32QAM). As such, small error in the phase states will affect 64QAM more than 16QAM.  This is indicative in the EVM data shown in Fig.~\ref{evmqam}. The point where 32QAM cannot be received (the right side of the EVM curve where the data stops) occurs at a smaller symbol rate than the point where 16QAM cannot be received, and 64QAM fall off even faster.

While BPSK and QPSK are pure phase modulation schemes, QAM requires modulation of both the phase and amplitude. The detected amplitudes from the atom-based mixer drops with higher IF values\cite{phase}, and it becomes hard to distinguish changes in amplitude (required for the QAM scheme). This explains why the QAM scheme degrades before BPSK and QPSK scheme.

The results in the paper illustrate the capability of using a Rydberg-atom mixer to detect and receive various phase and amplitude digital modulation schemes (BPSK, QPSK, 16QAM, 32QAM, and 64QAM).  The atom-based mixer can still detect and receive digital signals even when the transmitted symbol rate approached the bandwidth the Rydberg-atoms (around 1~MHz-10~MHz, which is the likely limit of the IF that can be used for the Rydberg atom-based mixer), keeping in mind that the EVM does increase with high symbol rate.  With that said, data can be received even for high EVM, however, error correction techniques may be required.  While the advantages of a Rydberg atom-based digital receiver have not been fully explored, the atom-based mixer potentially has many benefits over conventional technologies in detecting and receiving modulated signals.  For example, (1) no need for traditional demodulation/down-conversion electronics because the atoms automatically perform the demodulation for AM and FM signal\cite{rc3, amfmstereo} and automatically down-converts the phase modulated signals to an IF, (2) nano-size antennas and receivers over a frequency range of 100~MHz to 1~THz\cite{r2, r3}, (2) no Chu limit\cite{chu} requirements as is the case for standard antennas, (3) direct real-time read out,  (4) multi-band (or mutli-channel) operation in one compact vapor cell\cite{rc3, amfmstereo}, (5) the possibility of electromagnetic interference-free receiving, and (6) ultra-high sensitivity reception from 100~MHz to 1~THz\cite{weak}. Furthermore, there are indications that this Rydberg atom-based system  may be less susceptible to noise. As was the case in measuring CW electric-field strengths\cite{r20}, where we performed experiments measuring CW E-field strengths using this atom-based approach in the presence of band-limited white Gaussian noise and we showed that the E-field strength could be detected in low CW-signal to noise-power ratio conditions. The detection scheme discussed here can be improved by reducing laser noise and systematic effects, which is the topic of future work. While more research is needed to fully understand the advantages of the Rydberg atom approach over conventional radio technologies, the study reported here illustrates the capability of a Rydberg atom-based receiver/antenna system to detect and demodulate BPSK, QPSK, and QAM signals.
In effect, we are now in a position to be able to interrogate ensembles of atoms to such accuracy that we can use them to receive data from a transmitted communication signal.


{\bf Acknowledgment:} We thank Drs. R. Horansky (with the National Institute of Standards and Technology), and S. Voran (with the Institute for Telecommunication Sciences) for their useful technical discussions of phase modulated systems and the use of EVM.


\begin{thebibliography}{\footnotestlye}



\bibitem{r2} C.L. Holloway, M.T. Simons, J.A. Gordon, P.F. Wilson, C.M. Cooke, D.A. Anderson, and G. Raithel, {\it IEEE Trans. on Electromagnetic Compat.,} vol. 59, no. 2, 717-728, 2017.
\bibitem{r3} C.L. Holloway, J.A. Gordon, A. Schwarzkopf, D.A. Anderson, S.A. Miller, N. Thaicharoen, and G. Raithel, {\it IEEE Trans. on Antenna and Propag.,} vol. 62, no. 12, 6169-6182, 2014.
\bibitem{r4} J.A. Sedlacek, A. Schwettmann, H. Kubler, R. Low, T. Pfau and J.P. Shaffer, {\it Nature Phys.}, vol. 8, 819, 2012.
\bibitem{r5} C.L. Holloway, J.A. Gordon, A. Schwarzkopf, D.A. Anderson, S.A. Miller, N. Thaicharoen, and G. Raithel, {\it Applied Phys. Lett.,} vol. 104, 244102-1-5, 2014.
\bibitem{r6} J.A. Sedlacek, A. Schwettmann, H. Kubler, and J.P. Shaffer, {\it Phys. Rev. Lett.,} vol. 111, 063001, 2013.
\bibitem{r7} J. A. Gordon, C. L. Holloway, A. Schwarzkop, D. A. Anderson, S. Miller, N. Thaicharoen, G. Raithel, {\it Applied Physics Letters,} vol. 105, 024104, 2014.
\bibitem{r8} H. Fan, S. Kumar, J. Sedlacek, H. Kubler, S. Karimkashi and J.P Shaffer, {\it J. Phys. B: At. Mol. Opt. Phys.,} vol. 48, 202001, 2015.
\bibitem{r9} M. Tanasittikosol, J.D. Pritchard, D. Maxwell, A. Gauguet, K.J. Weatherill, R.M. Potvliege and C.S. Adams, {\it J. Phys B,} vol. 44, 184020, 2011.
\bibitem{r10} C.G. Wade, N. Sibalic, N.R. de Melo, J.M. Kondo, C.S. Adams, and K.J. Weatherill, {\it Nature Photonics}, vol. 11, 40-43, 2017.
\bibitem{r11} H. Fan, S. Kumar, J. Sedlacek, H. Kubler, S. Karimkashi and J.P Shaffer, {\it J. Phys. B: At. Mol. Opt. Phys.,} 48, 202001, 2015.
\bibitem{r12} D.A. Anderson, S.A. Miller, G. Raithel, J.A. Gordon, M.L. Butler, and C.L. Holloway, {\it Physical Review Applied,} 5, 034003, 2016.
\bibitem{r13} D.A. Anderson, S.A. Miller, A. Schwarzkopf, C.L. Holloway, J.A. Gordon, N. Thaicharoen, and G. Raithelet, {\it Physical Review A}, vol. 90, 043419, 2014.
\bibitem{r14} A.K. Mohapatra, T.R. Jackson, and C.S. Adams,  {\it Phys. Rev. Lett.,} vol. 98, 113003, 2007.
\bibitem{r15} C.L. Holloway, M.T. Simons, J.A. Gordon, A. Dienstfrey, D.A. Anderson, and G. Raithel, {\it J. of Applied Physics,} vol. 121, 233106-1-9, 2017.
\bibitem{r16} M.T. Simons,  J.A. Gordon,  and C.L. Holloway, {\it Applied Physics Letters}, vol. 108 174101, 2016.
\bibitem{r17} M.T. Simons,  J.A. Gordon,  and C.L. Holloway, {\it J. Appl. Phys.,} vol. 102, 123103, 2016.
\bibitem{r18} C.L. Holloway, M.T. Simons, M.D. Kautz, A.H. Haddab, J.A. Gordon, T.P. Crowley, {\it Applied Phys. Letters,} vol. 113, 094101, 2018.
\bibitem{r19} M.T. Simons, J.A. Gordon, and C.L. Holloway, {\it Applied Optics,} vol. 57, no. 22, pp. 6456-6460, 2018.
\bibitem{r20} M.T. Simons, M.D. Kautz, C.L. Holloway, D. A. Anderson, and G. Raithel, {\it J. of Applied Physics}, 123, 203105, 2018.

\bibitem{dan} D. Stack, B. Rodenburg, S. Pappas, W. Su, M. St. John, P. Kunz, M. Simons, J. Gordon, C.L. Holloway, ``Rydberg Dipole Antennas'', APS DAMOP, June 5-9, Sacramento, CA, May 2017.

\bibitem{rc1}  D.H. Meyer, K.C. Cox, F.K. Fatemi, and P.D. Kunz, {\it Appl. Phys. Lett.}, vol. 12, 211108, 2018.

\bibitem{rc2} K.C. Cox, D.H. Meyer, F.K. Fatemi, and P.D. Kunz, ``Quantum-Limited Atomic Receiver in the Electrically Small Regime'',  arXiv:1805.09808v2, June 19, 2018.

\bibitem{amfmstereo} C.L. Holloway, M.T. Simons, A.H. Haddab, J.A. Gordon, and S. Voran, ``A Multiple-Band Rydberg-Atom Based Receiver/Antenna: AM/FM Stereo Reception'', {\it IEEE Antenna. and Propog. Mag.}, 2019.

\bibitem{biterror}  M.T. Simons, A.H. Haddab, R. Horansky, and C.L. Holloway, ``Atom-based receiver: A study of bit-error for a pseudo-random bit stream in the presence of noise'', submitted to {\it Antenna and Wireless Porpagation Letter.} 2019.

\bibitem{rc3} D.A. Anderson, R.E. Sapiro, and G. Raithel, ``An atomic receiver for AM and FM radio communication'', arXiv:1808.08589v1, Aug. 26, 2018.

\bibitem{rc4} Z. Song, W. Zhang, H. Liu, X. Liu, H. Zou, J. Zhang, and J. Qu, ``The credibility of Rydberg atom based digital communication over a continuously tunable radio-frequency carrier'',  arXiv:1808.10839v2, Sept., 2018.

\bibitem{atomradio} ``Get ready for atom radio'', {\it MIT Technology Review}, June, 2018: https://www.technologyreview.com/s/611977/get-ready-for-atomic-radio/

\bibitem{phase} M.T. Simons, A.H. Haddab, J.A. Gordon, and C.L.~Holloway, {\it Applied Physics Letters}, {\bf 114}, 114101 2019.

\bibitem{weak}  J.A. Gordon, M.T. Simons, and C.L.~Holloway, ``Weak Electric-Field Detection with Sub-1 Hz Resolution at Radio
Frequencies Using A Rydberg Atom-Based Mixer,'' submitted to {\it Applied Physics Letters}, 2019.

\bibitem{qpsk} T.S. Rappaport, {\it Wireless Communications: Principles and Practice}. Prentice Hall PTR, Upper Saddle River, NJ, 1996, chapter 5.

\bibitem{evm} M. D. McKinley, K. A. Remley, M. Myslinski, J. S. Kenney, D. Schreurs, and B. Nauwelaers, ``EVM calculation for broadband modulated signals,'' in Proc. {\it 64th ARFTG Conf. Dig.,} pp. 45–62, 2004.

\bibitem{chu} L.J. Chu, {\it J. of Applied Physics,} vol. 19, 1163, 1948.

\end{thebibliography}
\end{document}